\documentstyle[12pt]{article}

         \def\be{\begin{equation}}
         \def\bea{\begin{eqnarray}}

         \def\ee{\end{equation}}
         \def\eea{\end{eqnarray}}
         \def\R{\rm {I\kern-.200em R}}
         \def\C{\rm {I\kern-.520em C}}
         \def\rd{{\rm d}}
         
         \def\c{\cite}
         \def\tr{{\rm tr}}

\begin{document}
\begin{titlepage}
\vspace*{5mm}
\begin{center} {\Large \bf Large--$N$ limit of the generalized 2D }\\
\vskip 1cm
{\Large \bf Yang--Mills theory on cylinder}\\
\vskip 1cm
M. Khorrami$ ^{a,c}$ \footnote {e-mail:mamwad@theory.ipm.ac.ir}
M. Alimohammadi$ ^{b}$ \footnote {e-mail:alimohmd@theory.ipm.ac.ir},

\vskip 1cm
{\it $^a$ Institute for Advanced Studies in Basic Sciences,}\\
{\it P.O.Box 159,Gava Zang,Zanjan 45195,Iran.}\\
{\it $^b$ Department of Physics, University of Tehran, North Karegar,} \\
{\it Tehran, Iran }\\
{\it $^c$ Institute for Studies in Theoretical Physics and Mathematics,}\\
{\it P.O.Box 19395-5746, Tehran, Iran}\\

\end{center}
\vskip 2cm
\begin{abstract}
Using the collective field theory approach of large--$N$ generalized
two--dimensional
Yang--Mills theory on cylinder, it is shown that the classical equation
of motion of collective field
is a generalized Hopf equation. Then, using the Itzykson--Zuber integral
at the large--$N$ limit, it is found that the classical Young tableau density,
which
satisfies the saddle--point equation and determines the large--$N$ limit
of free energy, is the inverse of the solution of this
generalized Hopf equation, at a certain point.
\end{abstract}
\begin{center}
PACS nubers: 11.10.Kk, 11.15.Pg, 11.15.Tk \\
Kewwords: large--$N$, Yang--Mills theory, generalized Hopf equation, density
function
\end{center}

\end{titlepage}
\newpage

\section{ Introduction}

The 2D Yang--Mills theory (YM$_2$) is a theoretical tool for
understanding one of the most important theories of particle physics, i.e.,
QCD$_4$. It is
known that the YM$_2$ theory is a solvable model, and in the recent years there
have
been much efforts to analyze the different aspects of this theory. The lattice
formulation of YM$_2$ has been known for a long time \cite{m}, and many of the
physical quantities of this model, e.g.
the partition function and the expectation
values of the Wilson loops, have been calculated in this context [2,3].
The continuum (path integral)
approach of YM$_2$ has also been studied in \cite{bl} and, using this approaches,
besides the above
mentioned quantities, the Green
functions of field strengths have also been calculated \c{ali1}.

It is known that the YM$_2$ theory is defined by the Lagrangian
$\tr(F^2)$ on a
Riemann surface. In an equivalent formulation,
one can use $i\tr(BF)+\tr(B^2)$ as the Lagrangian of this model, where
$B$ is an auxiliary
pseudo-scalar field in the adjoint representation of the gauge group. Path
integration over the field $B$ leaves an effective Lagrangian of the form
$\tr(F^2)$.

Now the YM$_2$ theory is essentially characterized by two important properties:
invariance under area-preserving
diffeomorphisms and the lack of propagating degrees of freedom.
These properties are not unique to the
$i\tr(BF)+\tr(B^2)$ Lagrangian, but rather are shared by a wide class
of theories, called the generalized 2D Yang--Mills theories (gYM$_2$'s).
These theories are defined by replacing the $\tr(B^2)$ term by an arbitrary
class function $f(B)$ \c{w2}. Several properties of gYM$_2$ theories have
been studied in recent years, for example the partition function [7,8],
and
the Green functions on arbitrary Riemann surface \c{mk}.

One of the important features of YM$_2$, and also gYM$_2$'s, is its behaviour
in the case of large gauge groups, i.e., the large--$N$ behaviour of $SU(N)$
(or $U(N)$) gauge theories. Study of the large--$N$ limit of these theories
is motivated on one hand by an attempt to find a string representation
of QCD in four dimension \c{dog2}.
It was shown that the coefficients of $1/N$ expansion of the partition
function of $SU(N)$ gauge theories are determined by a sum over maps from a
two-dimensional surface onto the two-dimensional target space. These kinds of
calculations have been done in \c{gr1} and \c{gr2} for YM$_2$ and in
\c{gan} for gYM$_2$.

On the other hand, the study of the large--$N$ limits is useful in exploring
more general properties of large--$N$ QCD. To do this, one must calculate,
for example,
the large--$N$ behaviour of the free energy of these theories. This is done
by replacing the sum over irreducible representations of $SU(N)$ (or $U(N)$),
appearing in the expressions of partition function, by a
path integral over continuous Young tableaus, and calculating the
area-dependence of the free energy from the saddle-point configuration.
In \c{rus2}, the logarithmic behaviour of the free energy of $U(N)$ YM$_2$ on
a sphere with
area $A<A_c=\pi^2$ has been obtained, and in \c{dog3} the authors have considered
the case $A>A_c$ and proved the existence of
a third--order phase transition in YM$_2$. A fact that has been known earlier
in the context of lattice formulation \c{gr3}. In the case of gYM$_2$ models,
the same transition has been shown for $f(B)=\tr (B^4)$ in \c{ali2} and for
$f(B)=\tr (B^6)$ and $f(B)=\tr (B^2)+g\tr (B^4)$ in \c{ali3}, all on the sphere.

Such kinds of investigations are much more involved in the cases of surfaces
with boundaries. This is because in these cases, the characters of the group
elements, which specify the boundary conditions, enter in the expressions
of the partition functions and this makes the saddle--point equations too
complicated. In \c{gr4} (see also \c{dad}), the authors considered the YM$_2$ theory on cylinder
and investigated its large--$N$ behaviour. If we denote the two circles forming
the boundaries of the cylinder by $C_1$ and $C_2$, then the boundary conditions
are specified by fixed holonomy matrices
$U_{C_1}=$Pexp$\oint_{C_1}A_\mu (x){\rm d}x^\mu$
and $U_{C_2}=$Pexp$\oint_{C_2}A_\mu (x){\rm d}x^\mu$. In the large--$N$ limit,
in which the
eigenvalues of these matrices become continuous, the eigenvalue
densities of $U_{C_1}$ and $U_{C_2}$ are denoted by $\sigma_1(\theta )$ and
$\sigma_2(\theta )$, respectively, where $\theta \in [0,2\pi ]$. Then it was shown
that
the free energy of YM$_2$ on cylinder, minus some known functions, satisfies a
Hamilton--Jacobi equation with a Hamiltonian describing a
fluid of a certain negative pressure \c{gr4}. The time coordinate of this system
is the area of the cylinder between one end and a loop
($0\leq t \leq A$), and its position coordinate is $\theta$,
and there are two boundary
conditions $\sigma (\theta )|_{t=0}=\sigma_1 (\theta )$ and
$\sigma (\theta )|_{t=A}=\sigma_2 (\theta )$. It is found that the classical
equation of motion of this fluid is the Hopf (or Burgers) equation. Further, it
was shown that the Young tableau density $\rho_c$, satisfying
the saddle--point equation (and therefore specifying the representation which has
the
dominant contribution in the partition function at large--$N$), satisfies
$\pi\rho_c(-\pi \sigma_0(\theta ))=\theta$. $\sigma_0(\theta )$ is
$\sigma (\theta ,t)$ at a time (area) $t$ at which the fluid is at rest.
When $U_{C_1}=U_{C_2}$,
$\sigma_0(\theta )$ is the solution of Hopf equation at $t=A/2$.
In the case of a disc,
$\sigma_2(\theta )=\delta (\theta )$, the authors have calculated the critical
area $A_c$ by using the results of the Itzykson--Zuber integral \c{mat} at
large--$N$ limit.

Studying the same problem for gYM$_2$ has begun in \c{gopa}, in the
context of master field formalism.
In this paper we study this problem, gYM$_2$ on cylinder, using the
above described technique.
The plan of the paper is as following. In section 2, by calculating the classical
Hamilton--Jacobi equation, we obtain the corresponding Hamiltonian for the
eigenvalue density for almost general $f(B)$ and find the classical equations
of motion. It is found that these equations are the generalized Hopf equation.
In section 3, we show that the Young tableau density $\rho_c$ is the inverse
function of the solution of the generalized Hopf equation at some certain time
(area).

\section{ The collective field theory and the generalized Hopf equation}

As it is shown in [7,8,9], the partition function of a gYM on a cylinder
is
\be\label{1}
Z=\sum_R\chi_R(U_1)\chi_R(U_2)e^{-A C(R)},
\ee
where $U_1$ and $U_2$ are Wilson loops corresponding to the boundaries of
the cylinder, the summation is over all irreducible representations of the
gauge group, $\chi_R$ is the trace of the representation $R$ of the group
element, and $C$ is a certain Casimir of the group, characterizing the
particular gYM$_2$ theory we are working with. For the gauge group U($N$), the
group we are working with, the representation $R$ is labeled by $N$ integers
$l_1$ to $l_N$, satisfying
\be
l_i<l_j,\qquad i<j.
\ee
The group element $U$ (an $N\times N$ unitary matrix) has $N$ eigenvalues
$e^{i\theta_1}$ to $e^{i\theta_N}$. The character $\chi_R(U)$ is then
\be\label{3}
\chi_R(U)={{{\rm det}\left\{e^{i l_j\theta_k}\right\}}\over
          {J\left\{e^{i\theta_k}\right\}}},
\ee
where
\be\label{4}
J\left\{e^{i\theta_k}\right\} =\prod_{j<k}
       \left( e^{i\theta_j}-e^{i\theta_k}\right).
\ee
The Casimir $C$ is a function of $l_i$'s. In its simplest form, $C$ has an
expression
\be\label{5}
C(R)=\sum_{i=1}^N c(l_i),
\ee
where $c$ is an arbitrary function. Here, we restrict ourselves to this form.

In the Large-$N$ limit, it is convenient to define a set of scaled
parameters $y_i$, instead of $l_i$'s:
\be\label{6}
y_i:={{l_i}\over N}-{1\over 2}.
\ee
In the same limit, also two density functions $\rho$ and $\sigma$,
corresponding to the distribution of $y_i$'s and $\theta_i$'s, respectively,
are defined:
\bea
\sigma(\theta)&:=&{1\over N}\sum_{j=1}^N\delta (\theta -\theta_j)\\
\rho(y)&:=&{1\over N}\sum_{j=1}^N\delta (y-y_j)
\eea

Inserting (\ref{3}), (\ref{4}), and (\ref{5}) in (\ref{1}), using (\ref{6}),
and making some obvious redefinition of the function $c$, one arrives at
\be\label{8}
Z=K\sum_R{{{\rm det}\left\{ e^{iNy_j\theta_k^1}\right\}
           {\rm det}\left\{ e^{iNy_j\theta_k^2}\right\}}\over
           {{\cal D}\{\theta_k^1\}{\cal D}\{\theta_k^2\}}}
           e^{-NA\sum_k g(y_k)}.
\ee
Here we have defined
\be
{\cal D}\{\theta_k\}:=\prod_{j<k}\sin{{\theta_j-\theta_k}\over 2},
\ee
and $K$ is an unimportant constant.

To proceed, we use the change of variable
\be\label{10}
\tau_k:=i\theta_k,
\ee
and rewrite (\ref{8}) as
\be\label{11}
Z=\tilde K\sum_R{{{\rm det}\left\{ e^{Ny_j\tau_k^1}\right\}
           {\rm det}\left\{ e^{Ny_j\tau_k^2}\right\}}\over
           {D\{\tau_k^1\}D\{\tau_k^2\}}}
           e^{-NA\sum_k g(y_k)},
\ee
where
\be
D\{\tau_k\}:=\prod_{j<k}\sinh{{\tau_j-\tau_k}\over 2}.
\ee
We can then use, along the line of \cite{gr4},
\be
Z=\tilde K e^{N^2F},
\ee
and differentiate it to obtain
\be\label{14}
-{{\partial F}\over{\partial A}}={1\over{D^1 Z}}{1\over N}\sum_k
   g\left({\partial\over{N\partial\tau_k}}\right) (D^1Z),
\ee
where $D^1$ is $D\{\tau^1_k\}$. The function $g$ is assumed to be a
polynomial. So, to calculate the right--hand side of (\ref{14}), let's first
calculate it for a monomial. We have
\bea\label{15}
{1\over{ND^1 Z}}\sum_k\left({\partial\over{N\partial\tau_k}}\right)^n(D^1Z)
  &=&{1\over{ND^1Z}}\sum_{k,m}{n\choose m}\left[\left({\partial\over
  {N\partial\tau_k}}\right)^mD^1\right]\cr
  &&\times\left({\partial\over{N\partial\tau_k}}\right)^{(n-m)}Z\cr
  &=&{1\over N}\sum_{k,m}{n\choose m}\left[{1\over{D^1}}\left({\partial\over
  {N\partial\tau_k}}\right)^mD^1\right]\cr
 && \times\left[\left(N{{\partial F}\over
  {\partial\tau_k}}\right)^{n-m}+O\left({1\over{N^2}}\right)\right].
\eea
In the large-$N$ limit, one can of course omit the $O(1/N^2)$ term (which
contains higher derivatives of $F$). In this limit, one must also note
the limiting behaviours
\be
{1\over N}\sum_k b_k\to\int\rd\tau\;\tilde\sigma (\tau) b(\tau),
\ee
and
\be
N{{\partial b}\over{\partial\tau_k}}\to{\partial\over{\partial\tau}}
\left[{{\delta b}\over{\delta\tilde\sigma (\tau)}}\right]
\vert_{\tau =\tau_k}.
\ee
Here we have used a density function $\tilde\sigma (\tau )$ instead of
$\sigma (\theta )$, corresponding to the change--of--variable (\ref{10}).

Returning to (\ref{15}), we define
\bea\label{18}
D_k&:=&{1\over N}{\partial\over{\partial\tau_k}}\ln\vert D^1\vert\cr
   &=&{1\over{2N}}\sum_{j\ne k}\coth{{\tau_k-\tau_j}\over 2}.
\eea
This remains finite, as $N$ tends to infinity. One then has
\bea\label{19}
{1\over{D^1}}\left({1\over N}{\partial\over{\partial\tau_k}}\right)^m
D^1&=&\left({1\over N}{\partial\over{\partial\tau_k}}+D_k\right)^m\cr
   &=&\sum_l{m\choose l}D_k^{m-l}\left({1\over N}
      {\partial\over{\partial\tau_k}}+D_k\right)^l_{\rm s}.
\eea
Here, the subscript s denotes that part of expression which
contains only the derivatives of $D_k$, not $D_k$ itself. The first equality
simply comes from $(D^1)^{-1}N^{-1}(\partial/\partial\tau_k)D^1=N^{-1}
(\partial/\partial\tau_k)+D_k$. To obtain the second equality, one may
consider a term with $l$ factors of $D_k$. There are $m!/[l!(m-l)!]$ ways to
choose $l$ factors of $D_k$ from $m$ factors persent. The other $D_k$'s,
either are differentiated or are not present.

It may seem that this s part vanishes at the large-$N$ limit, since it
contains factors of $1/N$. This is, however, not the case, since there are
singular terms in $D_k$ at $j\sim k$. To calculate the non--vanishing part
of this expression, let us define a generating function:
\bea
q_k(u)&:=&\sum_m{{u^m}\over{m!}}\left({1\over N}{\partial\over{\partial\tau_k}}
+D_k\right)^m\cr
    &=&e^{u\left({1\over N}{\partial\over{\partial\tau_k}}+D_k\right)}.
\eea
This is easily seen to be
\bea\label{21}
q_k(u)&=&{1\over{D^1}}e^{{u\over N}{\partial\over{\partial\tau_k}}}D^1\cr
      &=&{{D^1\left(\tau_k+{u\over N}\right)}\over{D^1(\tau_k)}}\cr
      &=&\prod_{j\ne k}{{\sinh\left({{\tau_k-\tau_j+u/N}\over 2}\right)}
         \over{\sinh\left({{\tau_k-\tau_j}\over 2}\right)}}.
\eea
The s--part of this expression is contained in terms with small values for
$\tau_k-\tau_j$. To obtain this, we use
\be
\tau_k-\tau_j\approx{{k-j}\over{N\tilde\sigma (\tau_k)}},
\ee
let $j$ run from $-\infty$ to $\infty$ (but $j\ne k$), and keep only the
leading terms in $\tau_k-\tau_j$. It is easily seen that if one uses this
prescription for $D_k$ itself, $D_k$ vanishes. So, using the
above--mentioned prescription in (\ref{21}) gives exactly $q_{k{\rm s}}(u)$,
the s--part of $q_k(u)$. That is,
\bea
q_{k{\rm s}}(u)&=&\prod_{j\ne k}\left[{{{{k-j}\over{N\tilde\sigma (\tau_k)}}
 +{u\over N}}\over{{k-j}\over{N\tilde\sigma (\tau_k)}}}\right]\cr
               &=&\prod_{j<k}\left\{ 1-\left[{{u\tilde\sigma (\tau_k)}\over
                  {k-j}}\right]^2\right\}\cr
               &=&\prod_{n=1}^\infty\left\{1-\left[{{u\tilde\sigma (\tau_k)}
               \over n}\right]^2\right\}.
\eea
So,
\be
q_{k{\rm s}}(u)={{\sin[\pi u\tilde\sigma (\tau_k)]}\over
                      {\pi u\tilde\sigma (\tau_k)}}.
\ee
Having found this, we return to (\ref{19}) and arrive at
\be\label{25}
{1\over{D^1}}\left({1\over N}{\partial\over{\partial\tau_k}}\right)^m D^1=
\sum_l{m\choose l}D_k^{m-l} a_l[\pi\tilde\sigma (\tau_k)]^l,
\ee
where the coefficients $a_l$ are defined through
\bea
{{\sin x}\over x}&=:&\sum_{l=0}^\infty{{a_l}\over{l!}}x^l\cr
                 &=&\sum_{l=0}^\infty{1\over{l!}}{{\cos (\pi l/2)}\over{l+1}} x^l.
\eea
Inserting (\ref{25}) in (\ref{15}), one obtains
\bea\label{27}
{1\over{ND^1Z}}\sum_k\left({1\over N}{\partial\over{\partial\tau_k}}\right)^n
(D^1Z)&=&{1\over N}\sum_{k,m,l}{n\choose m}{m\choose l}a_l
 [\pi\tilde\sigma (\tau_k)]^l D_k^{m-l}\left( N{{\partial F}\over
 {\partial\tau_k}}\right)^{n-m}\cr
 &=&{1\over N}\sum_{k,l}a_l[\pi\tilde\sigma (\tau_k)]^l{n\choose l}\sum_m
 {{n-l}\choose{m-l}}D_k^{m-l}\left( N{{\partial F}\over{\partial\tau_k}}
 \right)^{n-m}\cr
 &=&{1\over N}\sum_{k,l}a_l[\pi\tilde\sigma (\tau_k)]^l{n\choose l}
 \left( N{{\partial F}\over{\partial\tau_k}}+D_k\right)^{n-l}\cr
 &=&\sum_l a_l{n\choose l}\int\rd\tau\;\tilde\sigma (\tau)
 [\pi\tilde\sigma (\tau)]^l\left\{{\partial\over{\partial\tau}}\left[
 {{\delta S}\over{\delta\tilde\sigma (\tau)}}\right]\right\}^{n-l}.
\eea
In the last step, we have defined a function $S$ through
\be
\cases{N{\partial\over{\partial\tau_k}}S:=N{\partial\over{\partial\tau_k}}F+
       D_k\cr {{\partial S}\over{\partial A}}:={{\partial F}\over
       {\partial A}}\cr}.
\ee
Combining (\ref{14}) with (\ref{27}), we arrive at
\be\label{29}
-{{\partial S}\over{\partial A}}=\sum_{n,l}{n\choose l}a_l g_n\int\rd\tau\;
\tilde\sigma (\tau)[\pi\tilde\sigma (\tau)]^l\left\{
{\partial\over{\partial\tau}}\left[{{\delta S}\over
{\delta\tilde\sigma (\tau)}}\right]\right\}^{n-l},
\ee
where $g_n$'s are the coefficients of the Taylor--series expansion of $g$.
Considering $A$ as a time variable, (\ref{29}) can be regarded as the
Hamilton--Jacobi equation corresponding to the Hamiltonian
\be\label{30}
H=\sum_{l,n}{n\choose l}g_na_l\int\rd\tau\;\tilde\sigma (\tau)
[\pi\tilde\sigma (\tau)]^l\left[{{\partial\tilde\Pi (\tau)}\over
{\partial\tau}}\right]^{n-l},
\ee
where $\tilde\Pi$ is the momentum conjugate to $\tilde\sigma$.

The summations in (\ref{30}) are easily carried out to yield
\be\label{31}
H={1\over{2\pi i}}\int\rd\tau\left\{ G\left[ i\pi\tilde\sigma (\tau)+
{{\partial\tilde\Pi}\over{\partial\tau}}\right] -G\left[ -i\pi\tilde\sigma
(\tau)+{{\partial\tilde\Pi}\over{\partial\tau}}\right]\right\},
\ee
where $G$ is an integral of $g$:
\be
{{\rd G}\over{\rd x}}=g(x).
\ee
From (\ref{31}), one can obtain the equations of motion for $\tilde\sigma$
and $\tilde\Pi$:
\bea\label{33}
\dot{\tilde\sigma}&=&{{\delta H}\over{\delta\tilde\Pi}}\cr
&=&-{1\over{2\pi i}}{\partial\over{\partial\tau}}\left[
g\left( i\pi\tilde\sigma (\tau)+{{\partial\tilde\Pi}\over{\partial\tau}}
\right) -
g\left( -i\pi\tilde\sigma (\tau)+{{\partial\tilde\Pi}\over{\partial\tau}}
\right)\right] ,
\eea
and
\bea\label{34}
\dot{\tilde\Pi}&=&-{{\delta H}\over{\delta\tilde\sigma}}\cr
&=&-{{i\pi}\over{2\pi i}}\left[
g\left( i\pi\tilde\sigma (\tau)+{{\partial\tilde\Pi}\over{\partial\tau}}
\right) +
g\left( -i\pi\tilde\sigma (\tau)+{{\partial\tilde\Pi}\over{\partial\tau}}
\right)\right] .
\eea
Defining
\be
\tilde v:={{\partial\Pi}\over{\partial\tau}},
\ee
as a {\it velocity} field, in correspondence with what defined in [18,19],
one can combine (\ref{33}) and (\ref{34}) into a
generalized Hopf equation:
\be\label{37}
\dot{\tilde v}\pm i\pi\dot{\tilde\sigma}=-{\partial\over{\partial\tau}}
[g(\tilde v\pm i\pi\tilde\sigma )]
\ee
In the case of YM$_2$, where $g(y_k)={1\over 2}y_k^2$, this equation reduces to
Hopf equation found in \c{gr4}. In this case, when one of the boundaries shrinks,
so that one has a disc instead of a cylinder, that is
$\sigma_2(\theta )=\delta (\theta )$,
the Itzykson--Zuber integral can be used to obtain a solution for the
Hopf equation and from that the critical area of the disc has been obtained
\c{gr4}. In our problem, gYM$_2$, we do not know such an integral representation.

\section{The dominant representation}
It is shown in \cite{mat} that the character $\Xi_R(U)$, can be written
as
\be\label{38}
\chi_R(U)=e^{N^2\Xi[\rho,\tilde\sigma]},
\ee
where, for large $N$,
\be\label{39}
\Xi[\rho,\tilde\sigma]=\Sigma[\rho,\tilde\sigma]+{1\over 2}\int\rd y\; y^2
\rho (y)+B[\tilde\sigma].
\ee
Here $B$ is some functional of $\tilde\sigma$, and $\Sigma$ satisfies a
Hamilton--Jacobi equation
\be\label{40}
{{\partial\Sigma}\over{\partial t}}={1\over 2}\int\rd x\;\mu (x)\left[
\left({\partial\over{\partial x}}{{\delta\Sigma}\over{\delta\mu}}\right)^2
-{1\over 3}\pi^2\mu^2(x)\right],
\ee
in which the variable $\mu$ satisfies the boundary conditions
\be\label{41}
\mu(x,t=0)=\tilde\sigma(x),
\ee
and
\be\label{42}
\mu(x,t=1)=\rho(x),
\ee
Defining $V$ as the derivative of the momentum conjugate to $\mu$, as in the
previous section, it is seen that the Hamilton--Jacobi equation (\ref{40})
is equivalent to the following evolution equation for $\mu$ and $V$.
\be\label{43}
{{\partial\Phi}\over{\partial t}}-\Phi{{\partial\Phi}\over{\partial x}}=0,
\ee
where
\be\label{44}
\Phi:=V+i\pi\mu.
\ee
and $V={\partial\over{\partial y}}{{\delta\Sigma}\over{\delta\mu}}$.

Using (\ref{39}) in (\ref{1}) and (\ref{11}), we see that in the large--$N$
limit, the dominant representation satisfies the following saddle--point
equation

\be\label{45}
\sum_i{\partial\over{\partial y}}{{\delta\Xi_i}\over{\delta\rho(y)}}=Ag'(y),
\ee
or
\be\label{46}
\sum_i{\partial\over{\partial y}}{{\delta\Sigma_i}\over{\delta\rho(y)}}=Ag'(y)
-\sum_iy,
\ee
where the summation is over boundaries. Now recall (\ref{37}). Rewriting it as
\be\label{47}
\dot{\tilde f}+{\partial\over{\partial\tau}}g(\tilde f)=0,
\ee
where
\be\label{48}
\tilde f:=\tilde v-i\pi\tilde\sigma,
\ee
one can write an implicit solution to (\ref{47}) as
\be\label{49}
f(\tau ,b)=f\{\tau-(b-a)g'[f(\tau , b)], a\},
\ee
where $a$ and $b$ are two particular values of the {\it time} variable
(here, actually the area). The same thing can be done to solve (\ref{43}).
In fact, one can define two functions $H^+$ and $H^-$ as
\be\label{50}
H^+(x):=x-(t-T)\Phi(x,T),
\ee
and
\be\label{51}
H^-(x):=x+(t-T)\Phi(x,t),
\ee
and see that $\Phi$ is a solution to (\ref{43}) if these two functions are
inverses of each other, i.e. $H^-[H^+(x)]=x$ \c{mat}.

As an ansatz for $H^\pm$ (with $t=1$ and $T=0$), we take
\be\label{52}
H^-_i:=A_ig'+i\pi\rho,
\ee
and
\bea\label{53}
H^+_i&:=&\tilde v_i-i\pi\tilde\sigma_i\cr
     &=&\tilde f_i
\eea
where
\be\label{54}
\tilde f(x)=\tilde f_0\{x-A_ig'[\tilde f_i(x)]\}.
\ee
Here $A_i$ is the area between a curve for which $\tilde f$ is $\tilde f_0$
and the $i$-th boundary. The meaning of (\ref{52}) and (\ref{53}) is that we
are seeking a solution to (\ref{43}) with boundary conditions
\be\label{55}
V_i(x,1)=A_ig'(x)-x,
\ee
and
\be\label{56}
\mu(x,0)=\tilde\sigma(x).
\ee
For such a solution, (\ref{46}) is obviously satisfied.

It is now easily seen that
\be\label{57}
H_i^-[H_i^+(x)]=A_ig'[\tilde f_i(x)]+i\pi\rho (\tilde f_0\{ x-A_i
g'[\tilde f_i(x)]\}).
\ee
If $\tilde f_0(x)$ is the inverse of $i\pi\rho$,
\be\label{58}
i\pi\rho[\tilde f_0(x)]=x,
\ee
then we have
\be\label{59}
H_i^-[H_i^+(x)]=x,
\ee

So (\ref{46}) is satisfied if $\rho$ satisfies (\ref{58}). But $\tilde f_0$
is generally a complex function, whereas $\rho$ should be real. It is now
better to return to the earlier variables $\theta$ and $\sigma$. We have
\be\label{60}
i\tilde\sigma (i\theta)=\sigma(\theta).
\ee
So (\ref{58}) can be written as
\be\label{61}
\pi\rho[-i f_0(x)]=x,
\ee
where
\be\label{62}
f:=v-i\pi\sigma.
\ee
The argument of $\rho$ in (\ref{61}) is real if $v_0$ is zero. So, if there
exits a loop on the surface, for which the velocity field is zero, then
there exists a dominant representation $\rho$, satisfying
\be\label{63}
\pi\rho [-\pi\sigma_0(\theta)]=\theta.
\ee
This is the same as that obtained in \cite{gr4}. Note, however, that the
equation governing the evolution of $\sigma$, (\ref{37}), is different from
the corresponding equation in \cite{gr4}.

\vspace {1cm}

\noindent{\bf Acknowledgement}

\noindent M. Alimohammadi would like to thank the Institute for Studies in
Theoretical Physics and Mathematics and also the research council of the
University of Tehran, for their partial financial supports.

\end{document}